\documentstyle[prl,aps,multicol,epsf,epsfig]{revtex}
 \newcommand \be {\begin{equation}}
\newcommand \ee {\end{equation}}
 \newcommand \ba {\begin{eqnarray}}
\newcommand \ea {\end{eqnarray}}                        
\begin{document}
\tightenlines

\title{\bf Aging in K$_{1-x}$Li$_x$Ta0$_3$: a domain growth interpretation}
\author{F. Alberici-Kious$^1$, J.P. Bouchaud$^2$, L.F. Cugliandolo$^3$, P. Doussineau$^1$, A. Levelut$^1$}
\address{$^1$ Laboratoire des Milieux D\'esordonn\'es et H\'et\'erog\`enes,\\
Universit\'e P. et M. Curie, Case 78, 75 252 Paris Cedex 05, France$^*$\\ 
$^2$ Service de Physique de l'Etat Condens\'e,
CEA-Saclay, 91191 Gif-sur-Yvette CEDEX, France\\
$^3$ LPTENS, 24, rue Lhomond, F-75231 Paris Cedex 05, France\\
and LPTHE, Tour 24, 4, Place Jussieu, F-75231 Paris Cedex 05, France.}

\maketitle

\begin{abstract}
The aging behaviour of the a.c. susceptibility of randomly substituted K$_{1-x}$Li$_x$Ta0$_3$
crystals reveals marked differences with spin-glasses in that cooling rate effects are very important. The response to temperature steps (including temperature 
cycles) was carefully studied. A model based on thermally activated domain growth accounts for all the experimental results, provided one allows for a large distribution of pinning energies, in such a way that `slow' and `fast' domains coexist.  Interesting similarities with deeply supercooled liquids are underlined.
\end{abstract}

\vfill


\begin{multicols}{2} 

\narrowtext

The a.c. susceptibility of `glassy' materials becomes time dependent 
at low temperatures: this is the aging phenomenon \cite{review}. Such a
behaviour has
been observed on spin-glasses \cite{expsg,sitges}, on disordered dielectrics
such as ammonium perchlorate \cite{gilchrist} or K$_{1-x}$Li$_x$Ta0$_3$ ({\sc
klt}) 
(which is an orientational glass \cite{Kli,klilog}), and more recently on a
deeply
supercooled liquid (glycerol) \cite{Nagel}. All these systems share common 
properties, but also important differences. The prominent features are the
following:

$\bullet$ When the system is cooled to a certain temperature $T_1$ below the
glass transition 
$T_0$, 
its frequency dependent response (which in the present paper is the real part
of the dielectric 
constant $\epsilon'$) depends on the time $\Delta t$ since the quench. This
dependence can be parametrized as \cite{review}:
\be
\epsilon'(\omega,t) = \epsilon'_{\sc st}(\omega)  + f(\Delta t) g(\omega)
\ee
where $f(\Delta t)$ is a (slowly) decreasing function describing the aging part and 
$\epsilon'_{\sc st}(\omega)$ is the stationary part of the response. In spin
glasses, the 
functions $f(u)$ and $g(u)$ both behave similarly (as $u^{-b}$ \cite{Dean,sitges}),
which means that 
the aging 
part of the response obeys approximatively a simple $\omega \cdot \Delta t$ scaling
(but see \cite{sitges,review}). 
On {\sc klt} crystals, $g(\omega)$ is nearly independent of
frequency, at least for temperatures not too close to $T_0$, while $f(\Delta t)$
behaves as a power
law with a small exponent \cite{Kli} or a logarithm, see Fig. 1. The situation 
is intermediate for glycerol, where the variation of $g(u)$ is somewhat weaker
than that of $f(u)$ \cite{Nagel}.

$\bullet$ The stationary part of the susceptibility  is
nearly independent of the cooling rate ${\cal R}$ in spin glasses \cite{Eric}.
This is in striking contrast with the case of {\sc klt}, where $\epsilon'_{\sc
st}(\omega)$ markedly depends on the cooling rate (see Fig. 1), decreasing 
roughly as $\log {\cal R}$, as the cooling rate is reduced. In other words, the apparent 
asymptotic value of the dielectric constant depends on the history, and 
thus certainly cannot be associated to an `equilibrium' response function.
Several cooling histories were probed in ref. \cite{Kli}, with the conclusion
that it
is essentially the time spent around $T_0$ which determines the value of
$\epsilon'_{\sc st}(\omega)$.
\vskip 0.2cm

\begin{figure}
\centerline{\psfig{figure=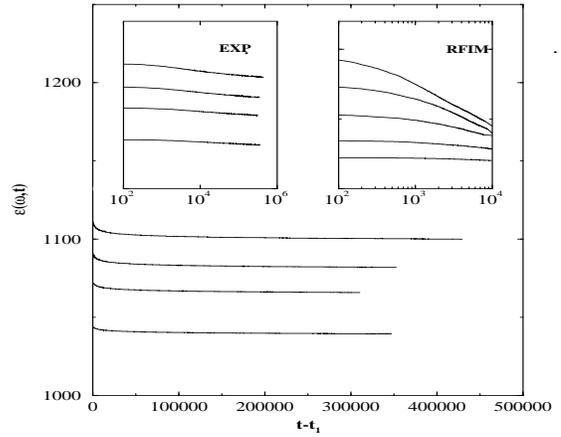,height=5cm,width=5.5cm}}

\vskip 0.5cm\caption{Plot of $\epsilon'(\omega/2 \pi=100\ KHz,t)$ for {\sc klt} ($x=0.025$) as a function of the time $\Delta t=t-t_1$ spent at temperature $T_1=4.2K$, for various cooling rates: ${\cal R}=0.045,\ 0.024,\ 0.011,\ 0.0058$ K/s. The slowest cooling rate corresponds to 
the lowest value of the apparent asymptote $\epsilon'_{\sc st}(\omega)$. Insets:
{\sc exp}: same data, in a log-linear scale; {\sc rfim}: numerical simulation
of the time evolution of the domain wall density in the 3-D Random Field Ising
Model, for different numerical cooling rates \protect\cite{footnote}.} \label{fig1}
\end{figure}

$\bullet$ Upon a negative temperature cycling $T_1 \to T_2 = T_1 - \Delta T
\to T_1$, the a.c. 
response function shows both {\it rejuvenation} and {\it memory}, in the
sense that new out of equilibrium processes are induced during the 
low temperature interval, but are erased once the temperature is back at
$T_1$:
dynamics resume as if the intermediate regime had not taken place. 
Memory is nearly perfect in spin-glasses when $\Delta T > 1K$, while in 
{\sc klt} (or glycerol) there appears an interesting `overshoot' -- see Fig. 2
below -- before the dynamics corresponding to the initial temperature can
resume. The
amplitude of this overshoot grows with the time $t_2$ spent at temperature
$T_2$.

The aim of this paper is to show that a picture based on domain growth, where
domain walls are pinned by static impurities, does account very well for all
the experimental data on {\sc klt}, provided one takes into account the fact
that the growth rate is actually very inhomogeneous, which leads to the
existence of both `slow' and `fast' domain walls. The relation with the
`droplet picture'
for spin-glasses will also be discussed in detail. Interestingly, the
similarities 
between {\sc klt} and glycerol suggest that some sort of pinned
domain growth, with a characteristic length scale, might also be present in
supercooled liquids.

When lithium is absent, KTa0$_3$ is known to be an `incipient' ferroelectric
at 
$T=0$, where quantum fluctuations prevent total ordering.
Thus, the ferroelectric correlation length $\xi(T)$ is expected to be large at
low temperatures: assuming that $\epsilon \propto \xi^2$, one finds that 
$\xi(T) \propto 1/\sqrt{T}$.
When lithium ions are randomly substituted, they create large 
local 
dipoles which are known to freeze individually around $T_0=40 K$, due to the
energy
barrier for dipole reorientation. Below this temperature, 
the lithium ions can thus be seen as giving rise to a random static electric
field, which act as a pinning field for the ferroelectric domain walls. 
The growth of the ferroelectric order towards its equilibrium value $\xi(T)$
is thus 
strongly impeded. The model we have in mind in thus the random field 
Ising model ({\sc rfim}) {\it above} its ferromagnetic transition, but close to it so that
the equilibrium domain size is somewhat larger than the lattice spacing $a$.
A numerical simulation
of the three dimensional {\sc rfim} with different cooling rates
actually reveals qualitatively similar features (Fig 1 -- inset) \cite{footnote}.

In the presence of pinning, several arguments \cite{RFIM,Nattermann} suggest that 
the typical energy barrier which impedes upon the motion of a domain of 
size $R$ is of the order of $E(R) \sim \Upsilon \left(\frac{R}{a}\right)^\theta$
where $\theta \simeq \frac{4}{3}$ for nearest neighbour interactions in three
dimensions and 
$\Upsilon$ is an energy scale which depends both on the random fields and on
the surface energy
of the domain walls. Note that dipolar fields are long range, so that $\theta$
might have a somewhat larger value ($\theta \sim 2$). The effect of pinning
becomes noticeable
when $E(R) \simeq kT$. For large times, the growth law is thus governed by
thermal activation $t \simeq \tau_0 \exp\left[\frac{E(R)}{T}\right]$, 
where $\tau_0$ is a microcopic time scale, and the Boltzmann constant $k$ has
been set to one. One thus obtains \cite{RFIM,Nattermann}:
\be
R(t) \simeq \left(\frac{T}{\Upsilon} \log (t + e^{\Upsilon R_0^\theta/T})
\right)^{\frac{1}
{\theta}} \qquad R_0 = R(t=0)\label{2}
\ee
where from now on we measure $t$ in units of $\tau_0$ and $R$ in units of $a$.
Eq. (\ref{2}) 
holds until the 
equilibrium size $\xi(T)$ is reached, beyond which domains no longer grow.
Taking the
derivative of Eq. (\ref{2}), one obtains the growth rate $\Gamma(t,T)$, 
in which the temperature can now be considered as time dependent, for example
as $T(t)=T_0-
{\cal R} t$. Noting that at $t=0$ domains are in equilibrium at $T_0$ (i.e.
$R_0=\xi(T_0)$), one can integrate back $\Gamma(t,T(t))$, and find the
domain size as
a function of time at temperature $T_1$. In order to compare with experimental
results, we make the natural assumption that the domain 
walls contribute to the dielectric susceptibility proportionally to their 
total surface. Per unit volume, this leads to an excess susceptibility given
by:
\be
\Delta \epsilon'(\omega,t) \propto \langle R(t)^{2}\rangle/\langle R(t)^{3}\rangle g(\omega)
\ee
where $g(\omega)$ is the (frequency dependent) wall mobility and $\langle ... \rangle$ denotes an average over all the domain sizes present in the sample.

Numerically, however, this simple model fails to reproduce 
the experimental results in the following sense:  $\Upsilon$ is either too
small, and one 
observes a decay of $\epsilon'$, which can be slow but towards an asymptotic value which is independent of the cooling rate, or too large, in which case $\epsilon'$ indeed strongly depends on
the cooling rate
but does not relax at all at $T_1$ (at least over the experimental time window). The
{\it simultaneous}
observation of the 
two effects actually requires some fast growing domains coexisting with slower
ones, 
corresponding to different local barrier heights $\Upsilon$. This is actually
expected: since
the pinning field is random, Eq. (\ref{2}) can only describe a typical
behaviour, with large
fluctuations corresponding to particularly efficient (or inefficient) pinning
regions. We shall 
thus assume that $\Upsilon$ has a rather broad distribution. Let us first
consider the situation at time $t_1$ when the temperature $T_1$ is
first reached (i.e. $T_0-{\cal R} t_1=T_1$). There, three types of domains can
be distinguished:

$\bullet$ `Fast' domains, which have been able to remain in equilibrium during
the whole 
cooling phase, and follow the evolution of the equilibrium length $\xi$. Their size is thus $\xi(T_1)$ and they no further evolve after
$t_1$. Imposing that $\Gamma(t,T) > d\xi/dt$, one can check that fast domains correspond to $T_p < \ell T_1^{1+\theta/2}/T_0^{\theta/2}$, where
$T_p=\Upsilon \xi(T_0)^\theta$ is the `initial' pinning energy and $\ell$ 
is approximately given by $\ln[T_1/{\cal R}\tau_0]$.

$\bullet$ `Slow' domains, which are in equilibrium in the first stages of the
cooling process, but fall out of equilibrium before the temperature reaches
$T_1$. This occurs when the domain growth 
rate $\Gamma(t,T)$ becomes less than $d\xi/dt$, which occurs at a temperature
$\left(T_p^2
T_0^\theta/\ell^2\right)^{1/(2+\theta)}$. This 
corresponds to $\ell T_1^{1+\theta/2}/T_0^{\theta/2} < T_p < \ell T_0$.

$\bullet$ `Frozen' domains, which fall out of equilibrium as soon as $T < T_0$
because of a large local pinning energy $T_p > \ell T_0$. Of course, these
domains are not really frozen but only evolve on extremely long time scales.

Collecting the contribution of these different domains to $\langle
R(t_1)^{2}\rangle/\langle R(t_1)^{3}\rangle$, and assuming that the scale
$T^*$ of the distribution of pinning energies $T_p$ is much larger than
$\ell T_0$, we find \cite{details} that the contribution of the walls to the
dielectric susceptibility is proportional to $1-\alpha \ell T_0/T^*$, where
$\alpha$ 
is a certain
function of $T_1/T_0$ \cite{details}. Hence, this model predicts that the excess dielectric susceptibility is linear in $\ln {\cal R}$, as indeed reported in
\cite{Kli}. The model also allows us to compute the subsequent evolution of $\epsilon$. Physically, after a time $t_1+\Delta t$, the `fastest' slow domains have managed to
reach $\xi(T_1)$, while the others 
slowly grow. Adding the different contributions, we find that, for large $\Delta t$
(more precisely for $\ell (T_1/T_0)^{(1-\theta)/2} \ll \ln \Delta t  \ll T^*/T_0$),
the dielectric constant decreases as $-\beta  T_0/T^* \ln \Delta t$, where $\beta$ is
another function of $T_1/T_0$ (but independent of ${\cal R}$). This logarithmic decay is again in good agreement with the experimental data (at least for large enough $\Delta t$): see Fig 1. Of course,
for exponentially long times, all domains (even the `frozen' ones) reach their
equilibrium size $\xi(T_1)$; the point is that the logarithm is such a slowly
varying function that, even on the rather long
time scales investigated here (3 $10^5$ seconds), the dielectric
susceptibility appears to 
asymptote a cooling rate dependent value $\epsilon'_{\sc st}(\omega)$.
\vskip 0.2cm
\begin{figure}
\centerline{\psfig{figure=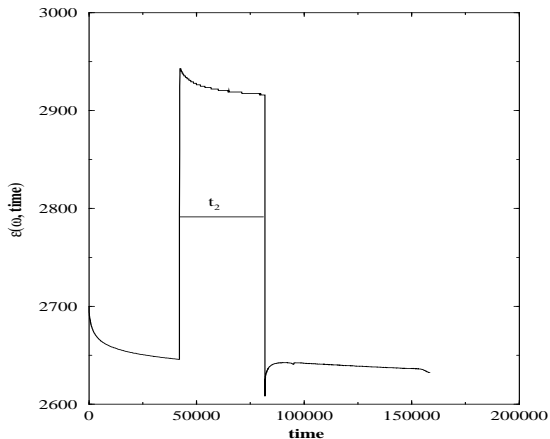,height=5cm,width=5.5cm}}

\vskip 0.5cm\caption{Effect of a negative temperature cycle on $\epsilon'(\omega/2\pi=100\ KHz,t)$ for {\sc klt} ($x=0.011$). The first period is spent at temperature
$T_1=18.4$ K during $40000$ s., then the sample is cooled to $4.2$ K for a time $t_2=40000$ s., and heated back again to $T_1$. The `overshoot' can be seen at
the beginning of the third period, where $\epsilon'(\omega,t)$ has a non monotonic behaviour.} \label{fig2}
\end{figure}

The interpretation of the cycling experiments, shown in
Fig 2, follows very similar lines. The main ingredient is again the existence
of the `fast' and `slow' domains
discussed above. When the temperature is reduced from
$T_1$ to $T_2$, the equilibrium length $\xi$ suddenly increases. Hence, the
`fast'
domains are driven out of equilibrium and must restart growing. Since the
temperature has
decreased, this is now a slow process, which accounts for the fact that
$\epsilon'$
decreases much like after the initial quench. (Note that the large
positive instantaneous change of $\epsilon'$ is due to the bulk contribution  
which is itself temperature dependent). The `slow' domains, which were already slow
at
$T_1$, are completely frozen at $T_2$ and hardly move.  Now, when the system is heated back 
to $T_1$, two things happen. First, the `fast' domains now have to shrink back to their 
equilibrium size. They have grown beyond $\xi(T_1)$ during their stay at lower temperature
 and consequently, the domain wall density is smaller than it should be. 
However, since the size of these domains is large, the barriers they have to jump are 
now high at $T_1$ and their shrinking is very slow. A more efficient process then occurs: 
internal nucleation of smaller domains. These nuclei grow according to Eq. (\ref{2}) from an 
initial size 
of the order of the lattice parameter $a$, until they reach the equilibrium 
domain size $\xi(T_1)$. This growth, rapid at the beginning, generates new
walls which induce an increase of the dielectric constant. This accounts for the `overshoot' observed just after the temperature step. Second, the `slow' 
domains which have been nearly unaffected by the temperature decrease, resume 
their dynamics in perfect continuity with the first period of time. Concerning the 
`overshoot', its amplitude ${\cal A}(t_2)$ depends on the time $t_2$ spent at low 
temperature, since
the number of net nucleation of domains is proportional to the volume occupied by
domains which have grown beyond $\xi(T_1)$. We thus expect ${\cal A}(t_2)$ to grow
slowly with time (since $R(t_2)$ only grows logarithmically) and then to saturate for
very long times. This is indeed what one observes experimentally. The overshoot contribution
can be parametrized as ${\cal A}(t_2)
 \exp -(\Delta t/\tau_1)$, where ${\cal A}(t_2)$ grows slowly with $t_2$, and 
{\it decreases} with $T_1$, a feature which our model also predicts \cite{details}.  $\tau_1$, on the other hand, only weakly depends on  $t_2$ or $T_1$ (see Figure 3).

\vskip 0.2cm

\begin{figure}
\centerline{\psfig{figure=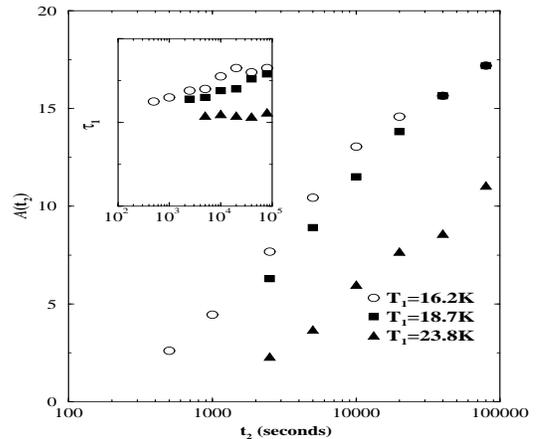,height=5cm,width=5.5cm}}

\vskip 0.5cm\caption{Amplitude of the overshoot ${\cal A}(t_2)$ as a function of the time
$t_2$ spent at $T_2=4.2$K, in log-linear scales, and for different temperatures $T_1$. Note that ${\cal A}$ increases roughly as $\ln t_2$ and decreases when $T_1$ increases. Inset: Dependence of the relaxation time of the overshoot $\tau_1$ on $t_2$. The $y$-axis scale is $0-2000$ seconds.} \label{fig3}
\end{figure}

Finally, we would like to discuss the differences between domain growth in a
random
field like system, which we argue to be a good model for {\sc klt} crystals,
and
`droplet' growth in spin-glasses \cite{BM,FH,KH}, which has been
advocated to be 
the relevant
picture (as opposed to models inspired from Parisi's mean field
`hierarchical' solution \cite{Dotsenko,LeFloch,Dean}, or exactly soluble dynamical mean-field
models \cite{KuCu,review}). First of all, domain growth cannot be trivial in spin glasses,
otherwise
strong cooling rate dependence would be seen, for example in the apparent
asymptote  of $\chi'(\omega,t)$, as is observed in {\sc klt} (see Fig. 1).
This is avoided by 
arguing that
the growing `phase' depends chaotically on temperature \cite{BM,FH,KH}, i.e. the structure of the equilibrium phase
changes altogether when temperature is varied. Hence,
the growth of the domains at temperature $T+\Delta T$ is useless to bring the
system
closer to equilibrium at $T$.
The
rejuvenation effect seen in {\sc klt} was attributed above to a change of the
{\it finite} equilibrium correlation length $\xi$ when the temperature is reduced, which
leads
to an overshoot effect not observed in spin-glasses. Conversely, in the droplet 
picture, the equilibrium size of the domains is infinite \cite{FH}. The interpretation of {\it rejuvenation} is then related to the fact
that 
domains of the new phase (i.e. the one stable at temperature $T_2$) do grow at
the expense of the old phase (the one stable at temperature $T_1$). However,
this must be also
compatible with the observed {\it perfect memory} upon reheating. How this is
possible is still very much a subject of debate
(see e.g. \cite{LeFloch,Eric}), but the conclusion that different
phases must somehow coexist (at least for finite times) appears difficult to
avoid.

To summarize, we have argued that the observed aging effects and strong
history dependence
of the dielectric susceptibility in {\sc klt} crystals could be understood in
terms of
slow, inhomogeneous ferroelectric domain growth in the presence of random
pinning fields. An
important aspect, needed to interpret the `overshoot' observed after a
positive
temperature jump, is that the equilibrium correlation length $\xi$ is finite.
It
would be interesting to compare our results with similar experiments on well
characterized random field systems, in particular in the ordered phase where
$\xi =\infty$. We have discussed the similarities and differences with other
systems, such
as spin-glasses or supercooled liquids. In particular, no overshoot has been
reported in spin glasses for large enough temperature jumps $T_2-T_1$, while
the data on
glycerol is quite similar to the one discussed here, and raises the
interesting 
possibility that some kind of domain growth, with a finite equilibrium size,
might also be relevant in supercooled liquids.

We thank J. Kurchan for having suggested the analogy with random field
systems and M. M\'ezard, R. Leheny, P. Nordblad and E. Vincent for many useful discussions.
\small{

$^*$ Associ\'e au C.N.R.S., U.M.R. 7603.

}

\end{multicols}

\begin{thebibliography}{99}
\bibitem{footnote} The simulation of the 3-D {\sc rfim} was performed on a lattice with $64^3$ sites. In units of the ferromagnetic coupling, the initial
temperature is $T_0=5$ and the final temperature is $T_1=0.57$. The random field
is $h=\pm 2$, and the cooling times are 10,20,40,80,160 MC sweeps per spin. Note
that we have simulated, for convenience, the {\sc rfim} below its ordering
temperature, instead of finite tuning the temperature to obtain a large (but finite) correlation length, as is the case for {\sc klt}.
\end{thebibliography}
\end{document}